\pdfoutput=1

\documentclass[twocolumn,prl,reprint,preprintnumbers,amsmath,amssymb,superscriptaddress,nofootinbib]{revtex4}

\usepackage{epsfig,psfrag,graphics,verbatim}
\usepackage{graphicx}
\usepackage{dcolumn}
\usepackage{bm}
\usepackage{color}

\newcommand{\liseven}{^{7}\textrm{Li}}
\newcommand{\lisix}{^{6}\textrm{Li}}
\newcommand{\beseven}{^{7}\textrm{Be}}

\newcommand{\hethree}{^{3}\textrm{He}}
\newcommand{\hefour}{^{4}\textrm{He}}

\begin{document}
\preprint{TUM-HEP 841/12}

\title{Lithium synthesis in microquasar accretion}

\author{Fabio Iocco}
\affiliation{The Oskar Klein Center for CosmoParticle Physics, Department of Physics, Stockholm University, Albanova, SE-10691 Stockholm, Sweden}

\author{Miguel Pato}
\affiliation{Physik-Department T30d, Technische Universit\"at M\"unchen, James-Franck-Stra$\beta$e, 85748 Garching, Germany}

\date{\today}

\begin{abstract}
We study the synthesis of lithium isotopes in the hot tori formed around stellar mass black holes by accretion of the companion star. We find that sizable amounts of both stable isotopes $\lisix$ and $\liseven$ can be produced, the exact figures varying with the characteristics of the torus and reaching as much as 10$^{-2}$ M$_\odot$ for each isotope. This mass output is enough to contaminate the entire Galaxy at a level comparable with the original, pre-galactic amount of lithium and to overcome other sources such as cosmic-ray spallation or stellar nucleosynthesis.
\end{abstract}

\maketitle

\section{Introduction} 
\par Lithium is one of the very few elements in our Universe whose abundance is significantly affected by big bang nucleosynthesis (BBN), nuclear spallation processes in cosmic rays, and stellar nucleosynthesis. In particular, its $\liseven$ isotope is observed in the atmosphere of relatively cold stars in our Galaxy, namely halo stars of varying metallicity and stars belonging to globular clusters. The very first observations \cite{Spite:1982} have revealed a remarkable flat trend in the abundance of $\liseven$ vs the metallicity of the target star, known as the ``plateau''; the most recent observations of halo stars (see \cite{Sbordone:2010zi} and references therein) show instead a clear ``meltdown'' of such single abundance value at the lowest metallicities, as well as a moderate, yet non-negligible dispersion at intermediate ones. The ``roof'' of this plateau still sits at a factor $\sim$3 below the prediction from standard BBN, a discrepancy that is known as the ``lithium problem''. If the value of the plateau is of primordial origin, the observations are almost impossible to reconcile with standard BBN, and exotic models are to be invoked \cite{Iocco:2008va, Jedamzik:2009uy}. If that value is the result of processes of stellar origin, it {\it is} possible to set up a stellar mechanism able to deplete the original abundance  \cite{Korn:2006tv}, but it is still left to demonstration how this mechanism can act over three orders of magnitude in the metallicity of the star, thus reproducing an almost flat envelope. Cosmic-ray spallations can also destroy $\liseven$, but it seems unlikely that they can reproduce the observed features \cite{Fields:2004ug}.

\par The aim of this Letter is not to propose a solution to the so-called ``lithium problem'', but if possible to add another piece to the puzzle, as our results go in the direction of harshening the discrepancy between the BBN value and the one observed in stars. This opens up a scenario that needs to be taken into account in future studies concerning this subject. The existence of stellar mass black holes accreting from companion stars is established through observations of so-called microquasars, visible in X-ray frequencies in our Galaxy \cite{Mirabel:1999fy}. We show that material accreting onto a stellar mass black hole from a companion star, in a disk with characteristics of a hot torus, can undergo nucleosynthesis, and produce sizable quantities of $\lisix$ and $\liseven$. The amount of both isotopes depends on parameters such as the mass of the black hole, the viscosity of the medium, the accretion rate and the expelled fraction, and it can reach up to 10$^{-2}$ M$_\odot$ within the allowed region of the parameter space.

\par We first describe the structure of accretion tori around microquasars, and then discuss our implementation of the nuclear reaction network and nucleosynthesis equations. We finally present our results, as well as draw some guidelines and suggestions for future work.

\section{Structure of hot accretion tori}\label{sectori}
\par Accretion onto black holes and neutron stars has long been recognised as a very efficient means of converting rest-mass into radiation. The simplest arguments are actually adequate to qualitatively account for the luminosities of compact X-ray sources (see e.g.~\cite{Remillard:2006fc} for a review); yet, self-consistent models are extremely complex and have been constructed in only a few limiting cases \cite{ShakuraSunyaev,NovikovThorne,Pringle81,ShapiroTeukolsky}. Keplerian accretion disks \cite{ShakuraSunyaev,NovikovThorne} are amongst the most widely-studied models. They form around the black hole in a binary as material from the companion star is captured and confined to almost-Keplerian orbits given its high angular momentum. The viscosity of the orbiting gas is supposed to play a central role in the accretion process, but its nature and magnitude are not well-understood. Ever since the seminal paper by Shakura \& Sunyaev \cite{ShakuraSunyaev}, it has been customary to parameterise the viscous stress in the azimuthal direction $f_\phi$ as a fraction of the gas pressure $P$, i.e.~$f_\phi=\alpha P$, where $\alpha\lesssim 1$ is an  unknown parameter assumed to be constant. The effect of viscous stress is two-fold: it removes the angular momentum at a slow but steady pace leading to a shrinking of gas orbits, and it generates (frictional) heat. In standard Keplerian disks, electrons and ions in the accreting gas are tightly coupled; since the former dissipate energy efficiently, the frictional heat induced by viscosity is readily radiated away giving rise to a thin and cool disk.

\par Another type of models, the so-called hot accretion tori, was originally proposed in \cite{Rees82} to account for radio jets in some active galactic nuclei. Consider a black hole of mass $M_{BH}$ accreting at a rate $\dot{M}$ through a disk of half-thickness $h(r)$. As in Keplerian disks, the momentum transfer is driven by a viscous stress of the form $f_\phi=\alpha P$, which leads to a radial inflow timescale $t_r(r)=\alpha^{-1}(r/h)^2(r/r_g)^{3/2}r_g/c$, where $r_g=G M_{BH}/c^2$. The core idea behind the hot torus model is to conceive an accretion configuration in which the inflow timescale $t_r$ is short enough as to prevent electrons and ions from coupling. This is achieved for low accretion rates, specifically \cite{Rees82} $\dot{m}\equiv \dot{M}c^2/L_{Edd} \lesssim 50 \alpha^2$, where $L_{Edd}=4\pi G M_{BH} m_p c/\sigma_T$ is the Eddington luminosity. In such regime the frictional heat is not efficiently radiated away but is stored, the accretion region is inflated to an ion pressure supported torus with $h(r)\simeq r$ and ions remain at the virial temperature, $k_B T = (r_g/r)m_p c^2/3 $. It is therefore not unnatural to reach ion temperatures of tens of MeV in the hot torus. Moreover, the conservation of mass combined with the inflow timescale defined above gives an estimate of the mean mass density of the gas in a cylindrical shell of radius $r$ and half-thickness $h\simeq r$ \cite{Jin90}: $\rho(r)=\alpha^{-1} \dot{m} (r/r_g)^{-3/2} 3m_p/(2\sigma_T r_g)$. Following \cite{Jin90} we take the hot torus to extend from $r_i=100 r_g$ down to $r_f=6 r_g$. Over this range electrons are relativistic \cite{Rees82} and we shall assume they are adiabatically heated inside $r<r_i$, $T_e(r)=T(r_i)(\rho(r)/\rho(r_i))^{1/3} < T(r)$. Notice that the hot torus is optically thin and not in local thermodynamic equilibrium -- the gas density is in fact lower and the temperature higher than in standard thin disks, suppressing free-free opacity. The main sources of photons in the torus are electron radiating processes. Although synchrotron and synchrotron self-Compton may play a role depending on the magnetic field and electron temperature, we shall assume bremsstrahlung is the dominant emission mechanism. We make nonetheless an explicit test of this hypothesis later on. The photon number density per unit energy in a cylindrical shell of radius $r$, $dn_\gamma/ dE_\gamma (E_\gamma,r)$, is computed by taking into account the bremsstrahlung emission in the whole torus (for simplicity we assume spherical symmetry in this step). This gives an estimate of the local photon spectrum, which will be used in the following.

\par It is not straightforward to estimate the fraction of microquasars that exhibit a hot torus and such estimate would probably span orders of magnitude. However, if the accreting material is heated up -- or supplied hot -- and is unable to radiate efficiently, a pressure-supported torus will naturally form. Since these requirements are astrophysically plausible \cite{Rees82}, we anticipate that a non-negligible fraction of microquasars will host hot tori.

\section{Nucleosynthesis in hot accretion tori}\label{secnucleosynthesis}
\par It is clear from the discussion above that extremely high temperatures are reached inside the torus and all sort of nuclear reactions can take place. In particular, reactions with threshold of tens of MeV -- unimportant in environments such as BBN or the cores of main sequence stars -- are easily triggered. This is for instance the case of $\alpha+\alpha$ reactions, especially important for the synthesis of light elements as pointed out in \cite{RamaduraiRees,Jin90,Ramadurai91}. In order to pinpoint all the potentially relevant reactions for the production or destruction of $\lisix$, $\liseven$ and $\beseven$, we make use of the network generator \url{NetGen} \cite{netgensite} and the NNDC database \cite{nndcsite}. Neglecting processes with at least one rare or fastly-decaying reactant or more than two reactants, we identify 74 distinct reactions of which 17 are photodissociations. For each reaction the $Q$-value and the threshold are computed with \url{QCalc} \cite{qcalcsite} while the cross-section as a function of the kinetic energy in the centre-of-mass frame $\sigma(T_{cm})$ is constructed using data from NNDC \cite{nndcsite} or parameterisations available in the literature \cite{Wagoner,Cyburt,SerpicoIocco} for specific cases. Whenever available, the astrophysical $S$ and $R$ factors are used as well. The rate of a given photodissociation is then 
\begin{equation}
\Gamma(r)=\int{dE_\gamma \, \frac{dn_\gamma}{dE_\gamma}(E_\gamma,r) \sigma(E_\gamma) c } \quad ,
\end{equation}
while for the remaining reactions the thermally averaged cross-section at temperature $T$ reads 
\begin{equation}
\langle \sigma v \rangle (T) = \int {dT_{cm} \, f(T_{cm},T) \sigma(T_{cm}) v(T_{cm})} \quad ,
\end{equation}
with $f$ the Maxwell-Boltzmann distribution, $v(T_{cm})\simeq (2T_{cm}/\mu)^{1/2}$ and $\mu$ the reduced mass of the reactants. We have cross-checked our $\langle \sigma v \rangle$ (where possible) against the parameterised results of \cite{Mercer,Cyburt,SerpicoIocco} as well as the publicly available \url{NACRE} database \cite{nacresite} and found good overall agreement for the temperatures of interest in those references, i.e.~$T\lesssim 10^9-10^{10}$ K.

\par We are now in position to write down and solve the Boltzmann equation for the abundances of $\lisix$, $\liseven$ and $\beseven$ in the hot accretion torus. 
At every time in our calculation we check that the abundances of the produced $\lisix$, $\liseven$ and $\beseven$ are small compared to those of the reactants, thus justifying the choice of not following self-consistently the abundance of the latter. Consider a cylindrical shell of radius $r$ and half-thickness $h\simeq r$, shrinking towards the black hole at a pace set by the inflow timescale $t_r$. In particular the time-radius conversion reads $t(r_i,r)=t_r(r_i)-t_r(r)$. The mean $\liseven$ number density in the shell, $n_{\liseven}$, obeys
\begin{multline}\label{eqBoltz}
\frac{dn_{\liseven}}{dt} = \sum_i{(1+\delta_{A_iB_i})^{-1} \langle \sigma v \rangle_i n_{A_i} n_{B_i}  } \\
- n_{\liseven} \left( \sum_j{\langle \sigma v \rangle_j n_{A_j}} + \sum_k{\Gamma_k} \right) \quad ,
\end{multline}
where the sums in $i$, $j$ and $k$ run over processes that produce $\liseven$, spallation reactions of $\liseven$ and photodissociations, respectively; $A_{i,j}$ and $B_i$ are generic reactants and $\delta_{A_iB_i}$ is unity in the case of identical reactants and zero otherwise. Analogous equations hold for $\lisix$ and $\beseven$. The reactant densities are assumed to track the gas density $\rho(r)$ at fixed abundances (see next Section for further details). For a given set $(M_{BH},\dot{M},\alpha)$ we solve the three coupled differential equations between $(r_i,t_i)=(100r_g,0)$ and $(r_f,t_f)=(6r_g,t(r_i,r_f))$ with initial conditions $n_{\liseven}=(Y_{\liseven}/m_{\liseven})\rho(r_i)$, where $Y_{\liseven}$ is the initial mass fraction of $\liseven$, and likewise for $\lisix$ and $\beseven$. We can then compute the mass of $\liseven$ in the (steady-state) accretion torus at any given moment, $M_{\liseven}^t=m_{\liseven} \int_{r_f}^{r_i}{dr \, 2\pi r \, 2h(r) \, n_{\liseven}(r)}$. After an interval of time $\Delta t$, the torus will have accreted a total mass $M_{acc}=\dot{M}\Delta t$ and will have expelled a $\liseven$ mass
\begin{equation}\label{Mli7}
M_{\liseven} = f_{exp} \, \frac{M_{\liseven}^t}{M_t} \, \dot{M}\Delta t \quad ,
\end{equation}
with $M_t$ the torus mass and $f_{exp}$ the fraction of expelled material. Analogous expressions hold for $\lisix$ and $\beseven$. We adopt $\Delta t=100$ Myr and $f_{exp}=0.5$ all along, but our results can be trivially scaled to other values. A value of order unity for the expelled fraction $f_{exp}$ has been frequently used \cite{Jin90,RamaduraiRees} as the presence of powerful jets -- both observationally confirmed and theoretically expected \cite{Rees82} -- can lead to significant ejection in hot tori.

\section{Results and discussion}\label{secresults}

\begin{figure*}[htp]
\centering
\includegraphics[width=0.4\textwidth,height=0.35\textwidth]{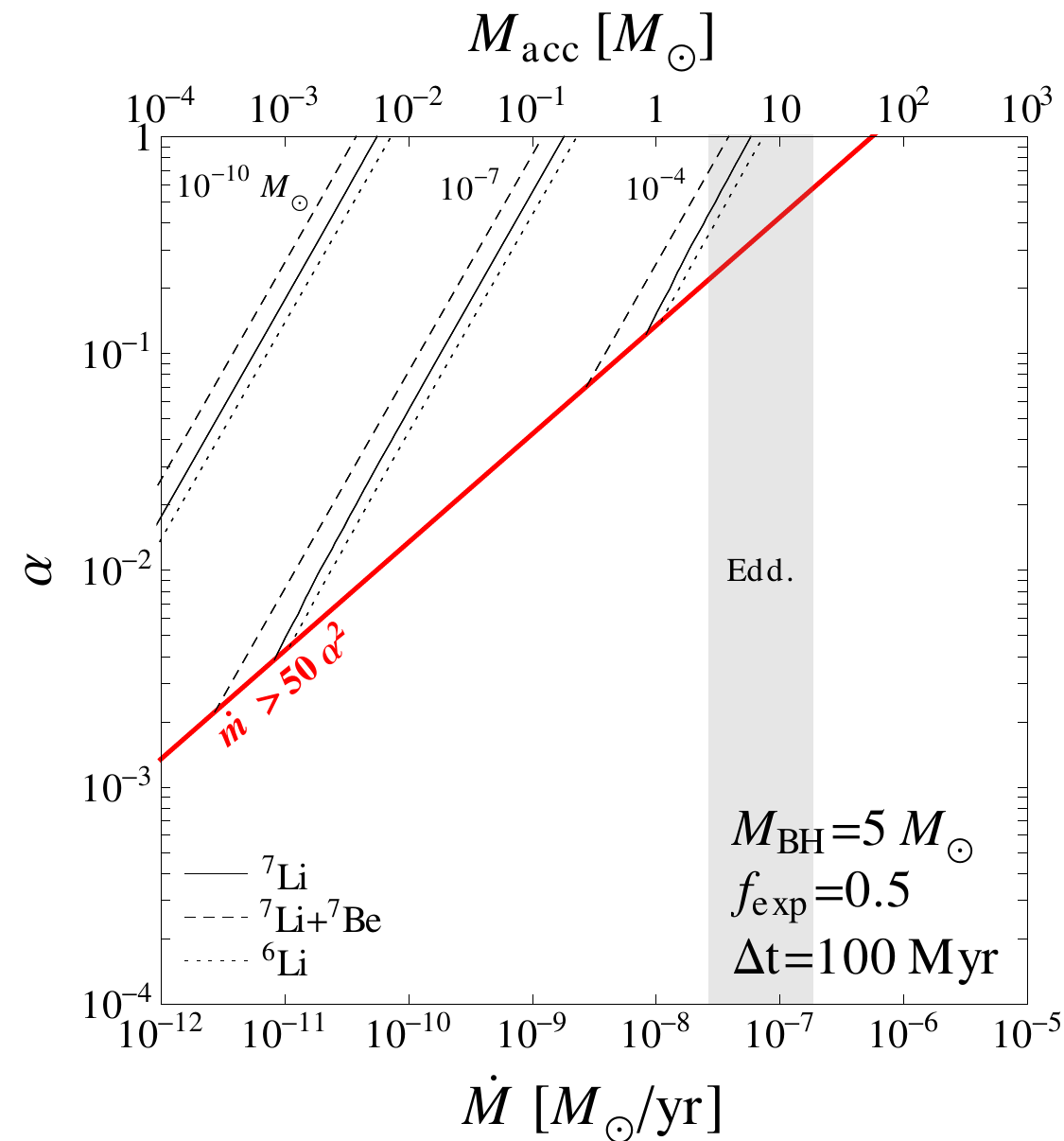} \hspace{0.5cm}
\includegraphics[width=0.4\textwidth,height=0.35\textwidth]{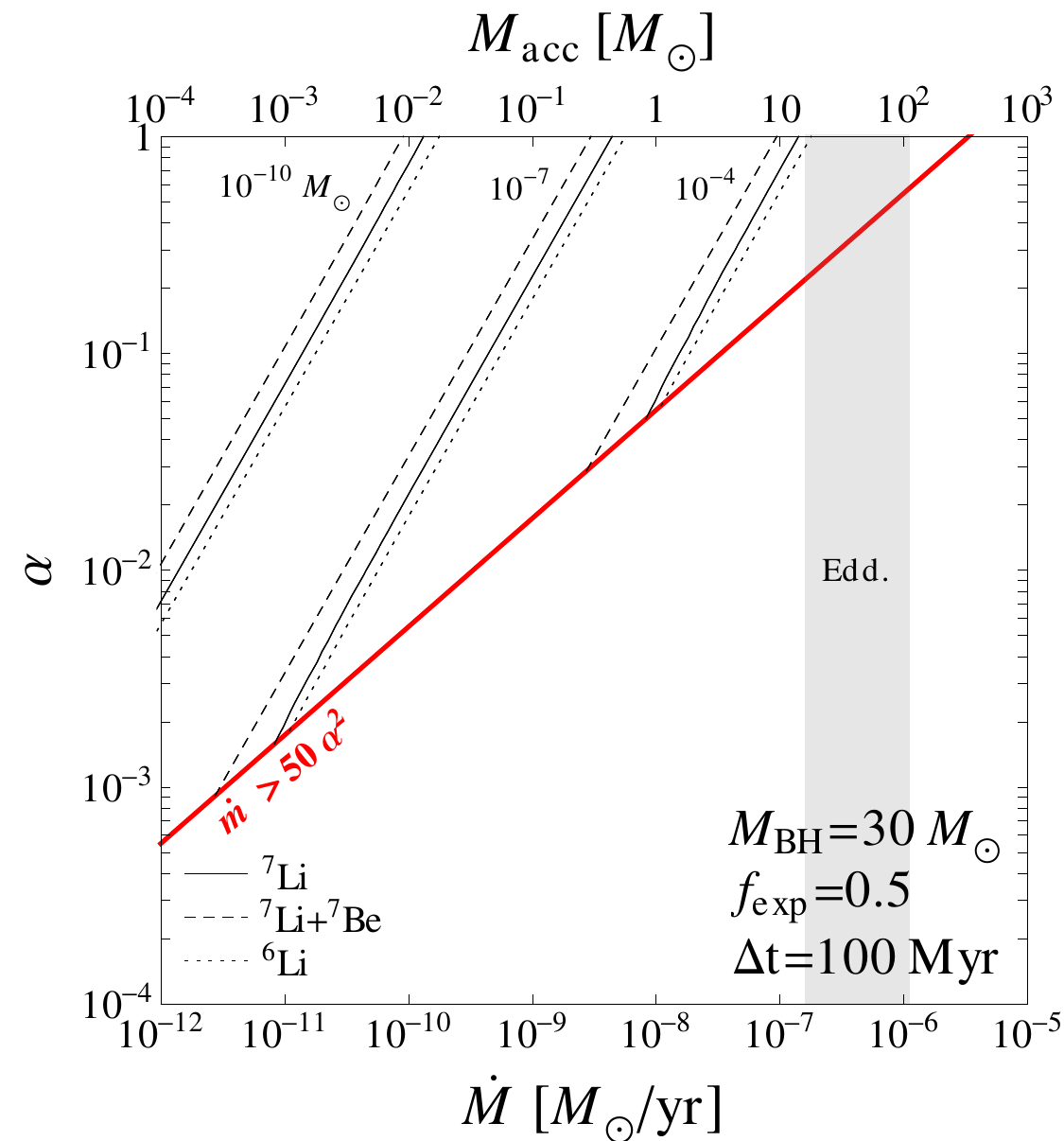}
\caption{The mass of $\liseven$ (plus $\beseven$) and $\lisix$ synthesised in a hot torus around a black hole of 5 M$_\odot$ (left) and 30 M$_\odot$ (right). The shadowed area encompasses the Eddington upper limit on the accretion rate $\dot{M}$ assuming accretion efficiencies $0.06\leq \epsilon\leq 0.42$.}
\label{fig:Mdotalpha}  
\end{figure*}

\par
\begin{figure}[htp]
\centering
\includegraphics[width=0.4\textwidth,height=0.35\textwidth]{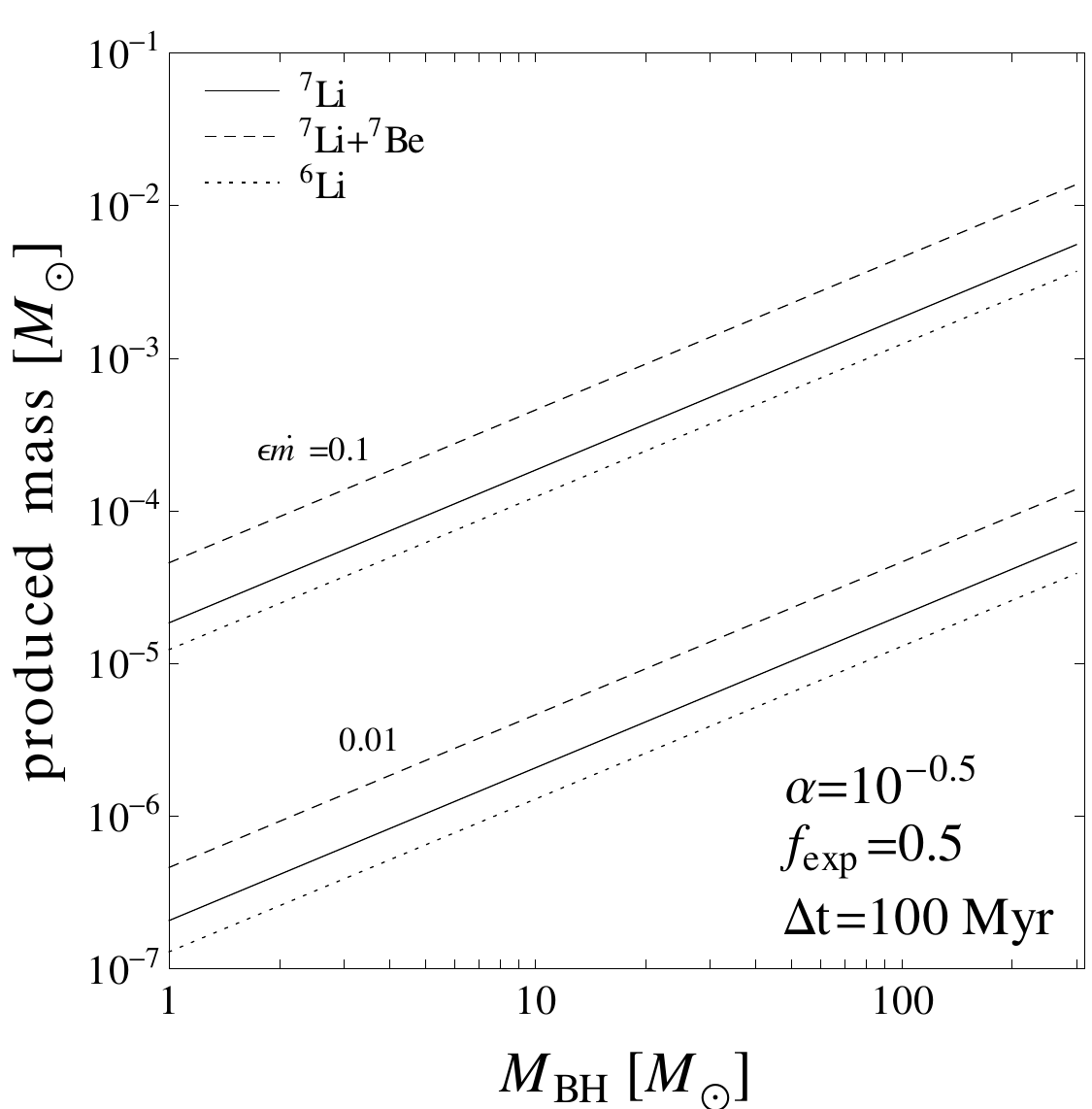}
\caption{The mass of $\liseven$ (plus $\beseven$) and $\lisix$ synthesised as a function of $M_{BH}$ for fixed $\alpha$ and $\dot{m}$. Here we take $\epsilon=0.06$.}
\label{fig:MBH}  
\end{figure}

 We have solved the set of coupled Boltzmann equations in a large region of the torus parameter space $(M_{BH},\dot M,\alpha)$. In our default case, we set the initial abundances of light elements to the ones leftover by BBN, namely \cite{Jedamzik:2009uy} $\textrm{D}/\textrm{H}=2.49\times10^{-5}$, $\hethree/\textrm{H}=10^{-5}$, $Y_P=0.2486$, $\liseven/\textrm{H}=5.24\times10^{-10}$ and $\lisix/\textrm{H}$=10$^{-14}$. However, we show later that the initial relative abundances are not crucial for the final amount of lithium isotopes synthesised. The results for sample black hole masses of 5 M$_\odot$ and 30 M$_\odot$ are shown in figure \ref{fig:Mdotalpha} on the plane $\alpha$ vs $\dot M$. It is clear that the nucleosynthesis in the hot torus can produce up to $\gtrsim10^{-4}$ M$_\odot$ of $\liseven$ (plus $\beseven$) and $\lisix$, with little dependence on the black hole mass (the small differences between the two cases arise mainly because of the different size of the torus and inflow time). On the other hand, the Eddington limit $L\equiv \epsilon \dot{M} c^2 \leq L_{Edd}$ -- indicated by the shadowed area in figure \ref{fig:Mdotalpha} with typical black hole accretion efficiencies \cite{ShapiroTeukolsky} $0.06\leq \epsilon\leq 0.42$ -- does depend on $M_{BH}$ and sets the most stringent constraint on the achievable amount of isotopes. We show in figure \ref{fig:MBH} the total produced mass of $\liseven$ (plus $\beseven$) and $\lisix$ as a function of $M_{BH}$ for fixed $\alpha$ and $\dot{m}$. Masses as large as 10$^{-2}$ M$_\odot$ are reached for hundred solar masses black holes. For the case of a 5 M$_\odot$ black hole (left panel in figure \ref{fig:Mdotalpha}), the lithium is produced at a ratio $^{6}\textrm{Li}/(^7\textrm{Li}+^7\textrm{Be})\simeq0.25-0.29$. Our results agree well with \cite{Jin90} when we adopt its equation (7) for the same black hole accretion configuration; however, in figures \ref{fig:Mdotalpha} and \ref{fig:MBH} we show lithium masses according to our equation \eqref{Mli7} that implicitly assumes a uniform ejection of torus material.  Computing the expelled mass as if it were ejected mostly from the inner regions yields final masses higher by a factor $\lesssim 10$, which would only strengthen our conclusions.

\par The nuclear reactions leading the synthesis of the lithium isotopes are the $\alpha+\alpha$ reactions: $\hefour(\alpha,n)\beseven$, $\hefour(\alpha,p)\liseven$, $\hefour(\alpha,d)\lisix$ and $\hefour(\alpha,np)\lisix$. All present similar thermally averaged cross-sections at the temperatures of interest and yield similar outputs of $\lisix$, $\liseven$ and $\beseven$ as shown in figure \ref{fig:Mdotalpha}. Therefore, the final abundance of lithium is sensitive mostly to the content of $\hefour$ present in the plasma. We find that in order to suppress lithium production by a factor $\sim$10$^4$ it is necessary to suppress the original abundance of $\hefour$ by a factor 10$^2$ (notice that the helium abundance enters quadratically in equation \eqref{eqBoltz}). Varying the initial abundances of D, $^3$H and $\hethree$ negligibly affects the final results, and adding an initial abundance of neutrons up to half of the number fraction has no significant effect. These checks show that the synthesis in the hot torus is sensitive mostly to its physical properties rather than the initial conditions of the gas chemistry (modulo the abundance of $\hefour$, which is the fundamental reactant for our purposes). It is worth stressing that synthesis at this level is possible because of the short resilience time of elements inside the hot torus. At photon densities below $\rho/m_p$, such as the ones considered here, the timescale for photodissociation of $\hefour$ in all possible channels (that require non-negligible amounts of photons above the threshold of $\sim$20 MeV) is much larger than the inflow timescale: for the most extreme cases we get $\Gamma^{-1}/t_r\gtrsim 3000$. Analogously, the photodissociation of $\lisix$, $\liseven$ and $\beseven$ is also found to be unimportant since $\Gamma^{-1}/t_r\gtrsim 70$. Together with the low lithium/beryllium densities, this prevents photodissociation from being competitive to the $\alpha+\alpha$ reactions. If we artificially modify the inflow time, we do in fact observe consistent depletion of lithium isotopes for longer inflow times, and enhanced final abundances for shorter resilience times. Furthermore, we find that scaling the photon density by a factor ranging from 10$^{-3}$ up to 10$^3$ has little effect on the lithium synthesis, whereas above 10$^4$ there are substantial variations. This large photon enhancement is rather unlikely if the photons are to be originated in the torus itself; however, if the cool disk surrounding the torus is nearby and produces large amounts of soft photons, these can be upscattered in the torus and induce the spallation of light elements as found in \cite{Mukhopadhyay:1999ut}.

\par The results in figures \ref{fig:Mdotalpha} and \ref{fig:MBH} are interesting under many regards. It is to be reminded that the amount of $\liseven$ and $\lisix$ produced by primordial nucleosynthesis, according to the most recent estimates of baryon density, are $\liseven/\textrm{H}=5.24\times10^{-10}$ and $\lisix/\textrm{H}=10^{-14}$; this represents the minimal ``background'' level of lithium isotopes which are present in the Galaxy at the time of its formation. A few simple estimates clarify the relevance of the results presented in this Letter: 
\begin{itemize}
\item It is expected that $10^8-10^9$ stellar mass black holes are present in our Galaxy, see e.g.~\cite{Remillard:2006fc} and references therein. The Milky Way mass in stars and gas is $M_{gal}\sim$ 5$\times$10$^{10}$ M$_\odot$ \cite{Gardner:2010sa}, which means that in principle, if a typical hot torus produces $10^{-4}\textrm{ M}_\odot$ of both $^7$Li and $^6$Li, then only 0.1$-$1\% ($10^{-6}-10^{-5}$\%) of the microquasars in our Galaxy need to host a hot torus so that the synthesised amount of $^7$Li ($^6$Li) is comparable to the BBN ``background'' level.
\item The first generation of stars (Population III, Pop.~III) is expected to form in small halos of total baryonic mass $10^5-10^6$ M$_\odot$. Recent numerical simulations of early star formation show evidence for multiple Pop.~III systems, with fragments of masses ranging in the sub-hundred solar masses regime \cite{Turk:2009ae, Smith:2011ac}. It has already been proposed that such systems may give rise to microquasars \cite{Mirabel:2011rx}: what we wish to highlight here is that the production of $10^{-4}-10^{-3}$ M$_\odot$ of $\liseven$ and $10^{-9}-10^{-8}$ M$_\odot$ of $\lisix$ by {\it a single} microquasar is enough to equate the BBN ``background'' level in the halo where the Pop.~III system has formed. These numbers are indeed achieved within the parameter space scanned in our study.
\end{itemize}

\par The previous examples are defective in some regards. In order for the quoted numbers to be taken at face value the material synthesised in the hot torus and expelled must be assumed to efficiently mix in the interstellar medium. This might be the case of primordial star-forming haloes -- where extremely efficient internal mixing is expected \cite{Scannapieco:2003rd} --, but not of the Milky Way, which raises the issue of even higher concentrations of lithium isotopes locally, around the microquasars themselves and in regions where microquasars are more frequent. Also, although it is not unreasonable that a significant part of the material gets expelled from the torus as argued above, presumably some of the expelled material would cross cool regions where significant spallation may occur. Finally, light element spallation within the torus can be important if a nearby, intense external source of photons is present. With these caveats in mind, and the need to explore further the physics of this ``aftermath'' part in a self-consistent way, it is clear however that the mechanism described has the potential to produce amounts of lithium comparable to the primordial values as well as to stellar nucleosynthesis and cosmic-ray synthesis, see discussion in e.g.~\cite{Fields:2004ug}. We point out that future population studies should take this mechanism into account, addressing especially the open issues discussed in this Letter. It is also worth remarking that microquasars are indeed observed in the X-ray bands in our Galaxy, and that in principle their optical and infrared counterparts are observable. This should give rise to the possibility of observing the amount of lithium (whose observable feature lies at 670.8 nm), thus checking the validity of the mechanism described so far or, alternatively, permitting to set constraints on the parameters and nature of the accretion onto stellar mass black holes. Interestingly, anomalously high lithium abundances have been observed in late type stars, companion of black holes or neutron stars (see \cite{Mukhopadhyay:1999ut} and references therein), thus suggesting another observational strategy to constrain the mechanism we have studied.


\par We have shown that nucleosynthesis of light elements taking place in the hot torus of accreting stellar mass black holes can produce sizable amounts of $\lisix$ and $\liseven$ for a wide range of parameters of the black hole and companion star system. Future studies concerning the ``lithium problem'' should take this mechanism into account, especially since its actual magnitude and impact can in principle be tested through observations.

\vspace{0.5cm}
{\it Acknowledgements } The authors would like to thank P.~D.~Serpico for useful comments. F.~I. is grateful to F.~Mirabel for having introduced him to the subject of microquasars, and thanks S.~P\'erez for useful references.


\bibliographystyle{apsrev.bst}
\bibliography{BHaccretionLi}

\end{document}